\documentclass[runningheads]{llncs}
\newcommand{\etal}{\textit{et al.}}
\newcommand{\circled}[1]{\raisebox{.5pt}{\textcircled{\raisebox{-.9pt} {#1}}}}
\usepackage{multirow}
\usepackage[T1]{fontenc}
\usepackage{graphicx}
\usepackage{hyperref}
\hypersetup{
  colorlinks=false,
  linkcolor=black,
  filecolor=black,   
  urlcolor=black,
  pdfborder={0 0 0}, 
}
\usepackage{xcolor}
\usepackage{footnote}
\makesavenoteenv{tabular}

\begin{document}

\title{\texorpdfstring{Where do developers admit their\\security-related concerns?}{}}

\author{Moritz Mock\inst{1}\orcidID{0009-0009-3156-6211} \and
Thomas Forrer\inst{2} \and
Barbara Russo\inst{1}\orcidID{0000-0003-3737-9264}}

\authorrunning{Mock \etal}

\institute{Free University Bozen-Bolzano,
Bolzano, Italy\\
\email{\{momock,brusso\}@unibz.it} \and
R\&D Department, W{\"u}rth Phoenix, Bolzano, Italy\\
\email{thomas.forrer@wuerth-phoenix.net}}

\maketitle       
\begin{abstract}
Developers use different means to document the security concerns of their code. 
Because of all of these opportunities, they may forget where the information is stored, or others may not be aware of it, and leave it unmaintained for so long that it becomes obsolete, if not useless. 
In this work, we analyzed different sources of code documentation from four large-scale, real-world, open-source projects in an industrial setting to understand where developers report their security concerns. In particular, we manually inspected 2.559 instances taken from source code comments, commit messages, and issue trackers. 
Overall, we found that developers prefer to document security concerns in source code comments and issue trackers. 
We also found that the longer the comments stay unfixed, the more likely they remain unfixed. Thus, to create awareness among developers, we implemented a pipeline to remind them about the introduction or removal of comments pointing to a security problem.
\end{abstract}

\setcounter{footnote}{0}
\section{Introduction}
Developers use different means to document their activity and store their artefacts. Generally speaking, source code is the preferred one, Kr{\"u}ger and Hebig \cite{Krueger2023}, but when specific tasks are concerned, the opportunities multiply. 
For instance, the most common places where developers document not-quite-right code that works (technical debt,~\cite{Cunningham92}) is again source code, but also commit messages, issue tracker, and pull requests, \cite{Li2023,Zampetti2021}. 

Source code comments are also used to store knowledge about security concerns,~\cite{Croft2022,Ferreyra2024}. \textit{Security Indicators} are keywords \cite{Croft2022} which are left behind by developers to express their worry connected to security within an application; this worry can be a vulnerability, potentially exploited by a third party.
Current studies also combine information from different sources to investigate the same concern. For instance, code and comments were explored to identify both technical debt and code vulnerabilities, \cite{Ferreyra2024,Russo2022}.
 
In this work, we aimed to answer the research question: \textit{Where do developers admit their
security-related concerns?}, therefore, we investigate the relevance of different sources where developers might express security concerns. The approach follows the idea that SAST/DAST tools, like SonarQube~\cite{SonarSource}, SemGrep~\cite{SemGrep} or Invicti~\cite{Invicti}, do not fully leverage textual data in the form of comments, commit messages, and issue trackers; hence we want to explore those sources to identify their usability.
We have mined three different sources (commit messages, comments, and issue tracker) from four large-scale projects from the industry with a history of up to 29k commits spanning 20 years. Furthermore, we have manually inspected 2.559 instances from different sources to understand which security indicators and sources are most relevant for developers.
We found that developers mostly use source code comments to admit security concerns; further investigations are needed to understand if there is a similar correlation between the severity of code, which is not quite right at the time of introduction, the longer it stays within the code base \cite{Bavota2016}. Additionally, we observed that those comments are either fixed soon (around ten releases) or they stay (almost) permanently. Therefore, we have developed a CI/CD bot that warns the developers whenever a security indicator is introduced or removed so that they are fully aware of comments that hint towards security concerns that are not captured by a SAST/DAST tool.
The replication package can be found here: \url{https://github.com/moritzmock/MiningSecurityIndicators}

\section{Methodology}

The following section describes some of the characteristics of the mined projects, how we performed the extraction of relevant information from the different sources for the commit messages, source code comments, and issue tracker and how the security indicators were evaluated against the inspected sources and repositories.

\textit{Characteristics of the mined projects - }
Table \ref{tab:characteristicsOfProjects} illustrates the characteristics of the four projects mined in the following section. It should be noted that the two projects are of the same origin, namely GLPI \cite{GLPI}; however, this was considered in the evaluation. Additionally, the number of \textit{years of history, \# releases, and \# commits} is limited as the project was created at the beginning of 2020. Furthermore, one of the evaluated projects, \textit{icingaweb2-module-slm}, is a project developed completely in-house, whereas the other projects are loaned from open source and specific extensions are elaborated for them.

\begin{table}[h!t]
  \centering
  \caption{Characteristics of the projects}
  \begin{tabular*}{\textwidth}{@{\extracolsep{\fill}}lrrrr@{}}
    \hline
    Project & years of history & \# releases & \# commits & lines of code \\ \hline
    GLPIv3 \cite{GLPI}&20.25&10&20.269&2.473.459 \\
    GLPIv4 \cite{GLPI}&4.25&34&195&2.557.746 \\
    icingaweb2 \cite{icingaweb2}&11&379&15.674&1.528.409 \\
    icingaweb2-module-slm& 4.83 &217&2.470&42.557 \\ 
    \hline
  \end{tabular*}
  \label{tab:characteristicsOfProjects}
\end{table}

\textit{commit messages - }
We leveraged PyDriller \cite{PyDriller}, which allowed us to easily access and extract the commit messages for each of the inspected projects. For the commit messages, we have mined all the commits and not only those of the release tags, as we did for the source code comments, else the inspection scope was too limited.

\textit{Source code comments - } We utilized PyDriller to extract the corresponding commit hash of the release tags, then checked out the repository at the obtained hash. At the given commit hash, we performed an analysis of all the present files, tracing each file with the corresponding comments in order to gather insights into how long it has been in the software repository and when it has been removed. We decided to inspect each tag rather than the commits to avoid including comments introduced for the short term, e.g., as a remark for the developer where they stopped and need to continue the next time.

\textit{Issue tracker (JIRA/Github) - } Some projects are loaned from open source, and others are developed internally. Therefore, two different issue trackers are used across the inspected projects. For JIRA, we used the Python package \texttt{jira} \cite{Jira} in combination with JQL-query (Jira Query Language to obtain all the relevant information. Whereas for GitHub, we used the available REST API~\cite{Github}.

\textit{Manual inspection - } The obtained data was analysed based on proportional sampling \cite{Demsar2006} with a confidence interval of 95\% and a marginal error of 5\% resulting values. Proportional sampling was selected as a methodology for reducing the overall sample size, which needs to be manually inspected. Resulting in 2.559 instances, 819, 1.076, and 683, respectively, for commit messages, source code comments, and issue trackers. The number of samples was calculated based on each individual source and summed up to final numbers. Additionally, we evaluated if pairs of security indicators occur and if those express a security concern only together or also individually.

\section{Results}
In the following section, each of the mined sources (commit messages, comments, and issue trackers) are discussed. In the following section, at the paragraph \textit{Manual Inspection}, some general observations are made regarding keywords that individually did not address any security concern but paired with others that were used by developers to address security concerns in source code.

\textit{commit messages - }
We observed that security indicators used in commit messages are used to address that certain parts of the code are now improved or fixed rather than admitting existing security issues in the code, e.g., ``fix search engine for XSS''. 
XSS (Cross-Site Scripting) is a security vulnerability in which a third party injects malicious scripts into a web page viewed by other users. This could potentially allow attackers to hijack user sessions or steal sensitive information. 
This observation is further supported by a quantitative evaluation of the data: 30\% of commits containing a security indicator have an associated issue linked to it, e.g., ``Fix minor bug in LDAP aliases \#2''.

\textit{Source code comments - }
We have analyzed the three sources (commit messages, source code comments, and issues tracker) and found that not all 288 security indicators are relevant to security-related issues. 
E.g., ``signature'', which was used for documenting how the \textit{digital signature} is handled within the application addressed by comments like: ``// signature widget''. Resulting in a different semantic for it compared to the originally intended one. 
Furthermore, we observed that some security indicators are removed after a relatively short time, usually around ten releases, or they stay permanently. Depending on the project \textit{years of history}, see Table \ref{tab:characteristicsOfProjects}, we observed a shift of the breaking point up to 200 release tags, which was the case for the project GLPIv3, which is maintained for more than 20 years and 20 thousand commits.
\begin{figure}[h!t]
  \centering
  \includegraphics[width=1\textwidth]{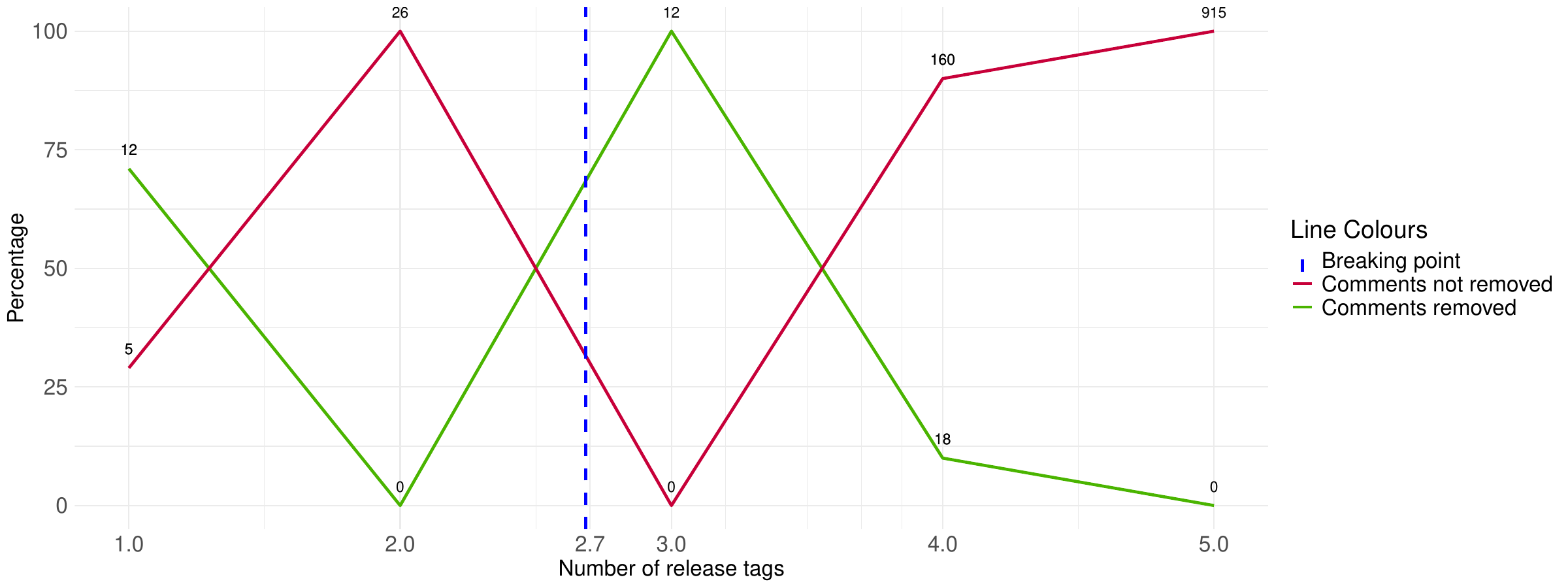}
   \caption{Percentages of comments that have or have not been removed over release tags for the project GLPI. After the breaking point of 2,7 release tags, the proportion of non-removed comments becomes greater. On top of the line, the absolute values can be found.}
  \label{fig:resultsManualInspection}
\end{figure}
Figure \ref{fig:resultsManualInspection} illustrates the number of release tags needed such that security indicators are removed after their introduction.
\textit{Issue tracker (JIRA/Github) - }
From the inspected instances, we observed that the security indicators like ``hack'', ``ldap'', and ``openssl'' are present due to sharing the same name with a PHP package rather than indicating a security concern. Whereas ``password'' and ``username'' are used in an example code to underline what the issue was; however, they were not related to a security issue.

\textit{Manual inspection - }
Our manual inspection of the different sources (commit message, comment, and issue tracker) classified 98 different security indicators, from which 79 were classified to be relevant. Additionally, we inspected whether security indicators appeared together and performed the inspection of their combination. 
In a work running in parallel \cite{Mock2024}, we performed an inspection of comments by three authors independently from each other for a manual classification of security indicators. 68.3\% of the keywords are identified in the work running in parallel and in this work as being relevant security indicators. The remaining 31.6\% are only identified by this work, which might be due to different sources, programming languages, temporal aspects, or sampling techniques that were leveraged in the two approaches. Inspecting those keywords that were only classified by this as relevant, many of them are related to login, e.g., ``two factor'', ``user account'', or ``user name'', hinting at the project-specific domains of user authentication, especially for GLPI~\cite{GLPI}.

At this point, we observed that quite general keywords such as ``ldap'' are used together with ``login'' or ``authentication'', which is not really surprising as LDAP (Lightweight Directory Access Protocol) is an authentication protocol. However, due to the connection with other keywords, more attention needs to be brought to seemingly irrelevant ones.
\section{Conclusion}
Mining various sources demonstrated the potential of combining them to increase the available context. We have manually inspected more than two thousand instances from different sources (commit message, comment, and issue tracker), unveiling that \circled{1} not every source has the same quality, i.e., in future work, we need to consider that not all sources provide the same information and richness, and \circled{2} not every indicator has a unique semantic meaning, making it harder to detect those keywords which in fact are indicating a security concern. To answer our initial question: \textit{Where do developers admit their security-related concerns?} Our preliminary results indicate that developers use source code comments as the most reliable means. The security indicators applied to the commit messages and issue tracker generate a lot of false positives, hence being a less reliable resource for identifying security admissions.

We plan to expand this work in future work by leveraging deep learning (DL) approaches to automate detection and classification.
Besides the challenges in designing a DL approach, building and maintaining trust in such an application is one, if not the major, issue we need to overcome. The DL is planned to be deployed in a CI/CD pipeline to automate the detection of different kinds of issues detached from the development process. 
Additionally, we plan to investigate the different behaviours of open source and industry developers, especially in a mixed state when practitioners from industry loan projects of open-source and expand them for their needs. Furthermore, how packages and projects are assessed and selected for internal use, i.e., which are the attributes developers consider most important to trust a third-party application.

\subsection*{Acknowledgements}
Moritz Mock is partially funded by the National Recovery and Resilience Plan (Piano Nazionale di Ripresa e Resilienza, PNRR - DM 117/2023). The research was carried out during an internship of Moritz Mock at W{\"u}rth Phoenix, Italy. The work has been funded by the project CyberSecurity Laboratory no. EFRE1039 under the 2023 EFRE/FESR program.
\bibliographystyle{splncs04}
\bibliography{ref}
\end{document}